\pdfoutput=1  %% arXiv admin added this line

\documentclass[aps,preprint,floats,epsf,epsfig,nofootinbib,letter]{revtex4-1}

\usepackage{amsmath}
\usepackage{amsfonts}
\usepackage{amssymb}
\usepackage{graphicx}
\begin{document}
\baselineskip=15pt \parskip=5pt

\vspace*{3em}

\title{Dark Matter Mass Constrained by the Relic Abundance, Direct Detections, and Colliders}

\author{Ho-Chin Tsai}
\email{tsaihochin@gmail.com}
\author{Kwei-Chou Yang}
\email{kcyang@cycu.edu.tw}

\affiliation{Department of Physics, Chung Yuan Christian University, Chung-Li 320, Taiwan}

\date{\today $\vphantom{\bigg|_{\bigg|}^|}$}

\begin{abstract}
We take into account a generic form of a Dirac fermionic dark matter (DM), which communicates with the Standard Model quarks via a  scalar mediator, in a model-independent way.  Four special interaction scenarios are investigated, where one is parity conserving and the other three are parity violating. Three of them result in the $v$ suppressed DM-nucleon cross sections, where $v \sim 10^{-3} c$ is the velocity of the DM in the laboratory frame. We constrain the masses of the dark matter and mediator as well as the
couplings from the thermal relic abundance, and the recent results of the XENON100 direct detection and collider experiments involving the two channels: (i) monojet plus large  missing transverse energy, and (ii) dijet. The current monojet  constraint is  not  stronger than that from the requirement of the correct relic density and the null result by the XENON100 direct detection.  We find that the dijet resonance measurements can exclude a large part of the parameter space $(m_\chi, m_Y)$, where the couplings for the mediator coupled to the dark matter and to the quarks are small and have roughly the same magnitude. The constraint from indirect detections  and diphoton resonance searches is also briefly discussed.
\end{abstract}
\maketitle
\newpage

\section{Introduction}
The nature of the dark matter is a challenging problem in the modern science. From astrophysical observations there are gravitational
sources which are invisible to us besides the visible stars and galaxies. So far, we know little about the invisible objects and call them to be the
dark matter  (DM), which could be massive to gravitate and non-baryonic to keep the evidence of big-bang nucleosynthesis (BBN) intact.
The DM is now believed to be responsible for $\sim$ 23\% energy density of our Universe, where the visible stars, galaxies and dim stellar gas
only contribute $\sim$ 4\% \cite{pdg}. From the viewpoint of the particle physics, the DM may be a particle which goes beyond the
standard model, and its identity remains elusive.  One of the authors (K.C.Y.) has investigated the possibility that the galactic dark matter exists in an scenario where the phantom field is responsible for the dark energy \cite{Li:2012zx}.

The relic abundance can be used to determine the interaction strength between the thermal DM and the Standard Model (SM) particles. A stronger interaction keeps the dark matter staying in the thermal equilibrium longer, so that the Boltzmann factor further suppresses its number density.   To reproduce the correct dark matter relic \cite{Komatsu:2010fb}, the governed Boltzmann's equation shows that the annihilation cross-section $\langle\sigma v\rangle$ of
the dark matter into SM particles is of order $3\times 10^{-26}$cm$^3/$sec which is about 1 picobarn$\cdot c$ (pb$\cdot c$) \cite{Bertone:2004pz}. This ${\cal O}(1)$  pb  cross-section interacting with SM particles infers that the dark matter could be produced in recent collider experiments.  If the mass of the DM is of order several hundred GeV, the DM can undergo a typical interaction with the electroweak
scale. This is the so-called weakly interacting massive particle (WIMP) miracle. Therefore it was suggested that the dark matter can couple to the Higgs boson in the effective theory below TeV scale and this kind of model is the so-called Higgs portal model \cite{He:2008qm,He:2009yd,Aoki:2009pf,Pospelov:2011yp,Nabeshima:2012pz,LopezHonorez:2012kv,Djouadi:2011aa,Bai:2012nv}.  Substituting  the annihilation  cross-section $\langle\sigma v \rangle$  obtained from the recent WMAP data into the formula of  the partial wave unitarity given in   \cite{Griest:1989wd}, the upper bound  on the mass of the thermal dark matter is approximately $34$~TeV. Some efforts are devoted to directly searching for WIMPs with masses of order  $\lesssim 10$~GeV \cite{Kang:2010mh,Gondolo:2011eq}.

 A number of underground experiments, {\it e.g.}, XENON, CDMS and DAMA/LIBRA, have being performed to detect the DM directly
scattered by the nuclei \cite{Aalseth:2011wp,Bernabei:2008yi,Ahmed:2010wy,Aprile:2012nq}. Although the controversial signals were detected
by DAMA/LIBRA and CoGeNT groups,  the null result has been reported by XENON100 and CDMS groups, respectively \cite{Farina:2011pw,Khlopov:2010pq}. The XENON100 data lead not only to the strongest limit so far for  constraining the DM-nucleon cross section to below $10^{-44}$
cm$^2$ for $m_\chi \sim 100$~GeV (and $10^{-43}$ cm$^2$ for $m_\chi \sim 1000$~GeV), but have also the constraints for cross section to
below $10^{-41}$ cm$^2$ in the low mass region around 10 GeV \cite{Aprile:2012nq}.

The DM may be produced at the hadronic colliders. Since the DM interacts weakly with SM particles, they can escape from the detector.
DM signals could be relevant to the processes for jets plus  large missing transverse energy ($\not\!\! E_T$) in the final states. The process with monojet + missing transverse energy $\not\!\! E_T$ final states has been
reported by CDF \cite{Aaltonen:2008hh}, CMS \cite{Chatrchyan:2011nd}, and ATLAS \cite{Aad:2011xw,Martinez:2012ie}, and is one of the
main channels  for the dark matter searches at the hadronic colliders.  Recently, ATLAS \cite{Martinez:2012ie} has analyzed monojets with
varying jet $p_T$ cuts using an integrated luminosity of 1.00 fb$^{-1}$. For all of these experiments, no obvious excess has been observed
compared with SM backgrounds.

On the other hand,  if the dark matter interacts with the hadron via a mediator,  it is possible to find out this mediator from the dijet mass spectrum at the hadron colliders, for which the searches by the CDF \cite{Aaltonen:2008dn} and D0 \cite{Abazov:2003tj} collaborations have used data from $p \bar{p}$ collisions at the Tevatron, while searches by the ATLAS \cite{Aad:2011fq} and CMS \cite{CMS:2012yf,Chatrchyan:2011ns} have used data from $ p p$ collisions at the LHC \cite{Harris:2011bh}. However, all measurements show no evidence for the new narrow resonance.

In this paper, we take into account in a model-independent way that a Dirac fermionic DM can couple to the SM quarks via a  scalar mediator, where the interaction can be parity conserving or violating.   Adopting this framework, we will constrain the masses of the dark matter and mediator as well as the
couplings from the thermal relic abundance, and the recent results of the XENON100 direct detection and colliders involving the monojet measurements with large  missing transverse energy  \cite{Aaltonen:2008hh,Chatrchyan:2011nd,Aad:2011xw,Martinez:2012ie} and dijet resonance searches \cite{Aaltonen:2008dn,CMS:2012yf,Chatrchyan:2011ns,Aad:2011fq}.  Some works for the interaction between the DM and SM particles via a neutral spin-1 mediator can be found in Refs. \cite{Accomando:2010fz,Gondolo:2011eq,An:2012va, Frandsen:2012rk, Barger:2012ey}

In the effective Hamiltonian approach with a contact interaction between the DM and SM quarks, which is just suitable for the heavy
mediator and has been discussed in the literature \cite{Cao:2009uw,Cheung:2012gi,Bai:2010hh,Kanemura:2010sh,fan,effective-1,effective-2,lhc-1,lhc-2,lhc-3,lhc-5,Haisch:2012kf,MarchRussell:2012hi}, only an  annihilation topology (usually the $s$-channel)  contributes to
the processes for the thermal relic abundance. However, for the general case, not only $s$-channel, but also the $u$- and $t$-channels may enter to
participate the interactions, where the $u$- and $t$-channels can be switched off when the DM mass is smaller than the mediator or the $s$-channel is
predominant if the coupling between the DM and mediator is much smaller than that between the quark and mediator. (See also Fig. \ref{fig:chichi2yy}.)

The mass of the mediator for the DM interacting with SM particles could be comparable with the energy scale of the colliders, so that the interactions can be resolved and the mediator is produced on shell. In this condition, the description of the effective contact operator is no longer valid. There are two possible ways to detect the relevant effect at the colliders. One is to measure the monojet plus missing transverse energy  in the final state, for which in addition to the SM background it is dominated by  $\chi  \bar\chi$  $+ {\rm monojet}$, with $\chi$ being the dark matter.  The other one is to search for the mediator $Y$ in the dijet mass spectrum. We will show that the LHC monojet constraints on the DM mass as well as  related parameters could be competitive with but not stronger than the dark matter direct detection at the present stage. For the dijet search, interchange with the mediator $Y$ in the $s$-channel interaction is predominant in the bump like component of the resonances, for which the upper limit is set by the collider measurements.

This paper is organized as follows. We begin in Sec.~\ref{sec:DM-model} with a generic form of a Lagrangian, which describes the Dirac fermonic DM interacting with the SM quarks via the scalar mediator.
In Sec. \ref{sec:analyses}, we perform a detailed numerical analysis for the relevant operators, and obtain the constraint on the parameter space of the DM and  mediator
masses. Together with discussions for the constraint from indirect detections and  diphoton resonance searches, we conclude in Sec. \ref{sec:conclusions}.

\section{The Dirac fermionic dark matter model}\label{sec:DM-model}

 In the present work,  we are devoted to the study of the fermionic dark matter, which interacts with the standard model quarks via a scalar  mediator, in a model-independent way. The relevant parts in the Lagrangian are
\begin{eqnarray}\label{eq:L}
\Delta{\cal L}=\overline{\chi} (i\not\!\partial-m_\chi) \chi+ \frac{1}{2} (\partial_\mu Y\partial^\mu Y-m_Y^2 Y^2)+  Y \overline{\chi}
(\lambda_s^\chi+i \lambda_p^\chi\gamma^5 ) \chi + Y \bar f (\lambda_s^f+ i \lambda_p^f\gamma^5) f \,,
\end{eqnarray}
where $f$ is a unspecified SM quark,  $\chi$ the fermionic DM, and $Y$  the scalar mediator which is the SM gauge singlet.  The
fermionic  DM that we consider  is assumed to be Dirac-like. (All the present calculations can equally apply  to the Majorana case with a
$Z_2$ parity to ensure its stability.)  The fermionic DM interacting with the SM particles through the scalar mediator is renormalizable and its origin is model-dependent. In this paper, we consider the case for the exotic $Y$ which couples to quarks with  the universal coupling, i.e., values of  $\lambda_{s}^f$ and $\lambda_{p}^f$ are independent of the flavor of the quarks, $f$.

On the other hand, in a general case, $Y$ could be the Higgs-like particle. The Higgs portal model is a nature generalization to connect the DM and SM sectors as studied in the literature.  Our effective  Lagrangian can be in connection with the Higgs portal model of the fermionic DM under some typical conditions.  For instance, we take the following potential form of the Higgs portal model as an example,
 \cite{LopezHonorez:2012kv,Pospelov:2011yp}\footnote{
The effective portal model
\begin{eqnarray} \label{eq:higgs2}
\Delta{\cal L}= - m_0 \overline{\chi}  \chi + \frac{ \lambda_{h\chi\chi}}{\Lambda_s} H^\dagger H \bar\chi \chi \,,
\end{eqnarray}
was adopted in \cite{Kanemura:2010sh,Djouadi:2011aa}.
Using the substitution in Eq. (\ref{eq:L}): $Y\to  h$, $\lambda_s^f \to -m_f/v$, $\lambda_p^f =\lambda_{p}^\chi=0 $,  $\lambda_{s}^\chi
\to \lambda_{h\chi\chi} v/\Lambda_{s}$, and $m_\chi \to \mu_0 - \lambda_{h\chi\chi} v^2/(2\Lambda_{s})$, one can relate our result to this Higgs portal model .}
\begin{eqnarray}
\Delta{\cal L} &=& - \mu_\chi \overline{\chi}  \chi +  \varphi \overline{\chi} g_s^\chi \chi  -\mu_H^2 H^\dagger H + \lambda (H^\dagger H)^2  -\frac{\mu_\varphi^2} {2} \varphi^2 \nonumber\\
&& + \frac{\lambda_\varphi}{4} \varphi^4  +
\frac{\lambda_4}{2}  \varphi^2 H^\dagger H +\frac{ \mu_1^3}{\sqrt{2}}\varphi +  \frac{ \mu_3}{2\sqrt{2}}\varphi^3+ \frac{\mu}{\sqrt{2}}\varphi (H^\dagger H) \,,
\end{eqnarray}
where $H$ is the SM Higgs doublet, for which the neutral component is shifted to $(v+h)/\sqrt{2}$ with $v=246$~GeV, and $\varphi=\phi+v_2$ is
the real singlet scalar with the spontaneously  breaking vacuum $v_2$.  The terms involving $\lambda_4$ and $\mu$ offer the Higgs portal
between the dark and SM sectors.  The two mass eigenstates correspond to the following superposition,
\begin{eqnarray}
\left(
\begin{array}{c}
 H_1  \\
 H_2
\end{array}
\right)
=\left(
\begin{array}{cc}
\cos\alpha  &   \sin\alpha \\
-\sin\alpha  & \cos\alpha
\end{array}
\right)
\left(
\begin{array}{c}
h  \\
 \phi
\end{array}
\right) \,,
\end{eqnarray}
where the mixing angle $\alpha$  satisfies
\begin{eqnarray}
\tan2\alpha=\frac{2\lambda_4 v v_2 + \sqrt{2} \mu v}{2 \lambda_H v^2 -2 \lambda_\phi v_2^2 + (2\mu_1^3 + \mu v^2)/(2\sqrt{2} v_2)}\,.
\end{eqnarray}
We adopt the convention $m_{H_1} < m_{H_2}$ and identify $H_1\approx h $ to be the SM-like Higgs for a small $\alpha$.  Comparing this Higgs portal model with our generic form of the DM, we have the following correspondences.  (i) If taking $Y$ to be $H_1$, which will be the SM-like Higgs as for $\alpha\to 0$, and adopting the limit $m_\chi \ll m_{H_2}$,  we have $ \lambda_s^\chi \simeq g_s \cos\alpha (m_{H_2}^2-m_{H_1}^2)/m_{H_2}^2$, $\lambda_s^f=-\sin\alpha \, m_f/v$ and  $\lambda_p^\chi=\lambda_p^f=0$. (ii) If taking $Y$ to be the heavier one $H_2$ and using the limit $m_\chi \ll m_{H_1}$,  we have  $ \lambda_s^\chi \simeq g_s \sin\alpha (m_{H_2}^2-m_{H_1}^2)/m_{H_1}^2$, $\lambda_s^f=-\cos\alpha \, m_f/v$ and  $\lambda_p^\chi=\lambda_p^f=0$. Note that in the multi-Higgs doublet or next-to-minimal supersymmetric
standard model (NMSSM) model \cite{Bird:2006jd}, $\lambda_p^f$ can be non-zero, although  it vanishes in the SM.

For the Higgs portal model, the solution with small coupling constants (corresponding to small $\alpha$), existing in the resonant region around  $m_\chi \approx m_{H_1}/2 \approx m_h/2$ or $m_{H_2}/2$,  can satisfy both the correct relic abundance and the XENON100 direct detection bound. Moreover, for this solution the allowed  invisible branching of the SM-like Higgs is  $\lesssim20\%$ \cite{LopezHonorez:2012kv}. A similar result for the Higgs portal model was obtained in \cite{Djouadi:2011aa}. The Tevatron and recent LHC data hint at that the SM-like Higgs may exist within  the very narrow window $m_h\sim 125$ GeV \cite{ATLAS:2012ae,Chatrchyan:2012tx,:2012gk,:2012gu,:2012zzl}.  If it is true, the discovery of Higgs can claim the establishment of the SM of particle physics.  The global fit for the  invisible branching of the SM Higgs can be found $e. g.$ in \cite{Espinosa:2012vu}.

For comparison, see also results for the $SS$ scenario given in Figs. \ref{fig:sigma_SI} and \ref{fig:sigma-mchi}, where, instead of $\lambda_s^f \propto m_f/v$, we have assumed the same value for $\lambda_s^f$ corresponding to an exotic $Y$.

\section{Numerical analyses}\label{sec:analyses}

\subsection{Dark matter constraint from the correct relic density}

The thermal relic abundance and freeze-out temperature are approximately given by \cite{KTbook}
\begin{eqnarray}
 \Omega_\chi h^2 \simeq \frac{1.07 \times 10^9 x_f}{\sqrt{g_*} m_{\rm pl} \langle \sigma_{\rm ann}
 v_{\rm rel}\rangle}, \qquad
 x_f\simeq \ln \frac{0.038 g_\chi m_{\rm pl} m_{\chi} \langle \sigma_{\rm ann} v_{\rm rel}\rangle}{\sqrt{g_*  x_f}},
 \end{eqnarray}
where $h$ is the Hubble constant in units of 100 km/(s$\cdot$Mpc), $m_{\rm Pl} = 1.22 \times10^{19}$ GeV is the Planck mass, $m_\chi$ is
the dark matter mass, $x_f = m_\chi/T_f$ with $T_f$ being the freeze-out temperature, $g_*$ is the number of relativistic degrees of freedom
with masses less than $T_f$, $g_\chi$ is the number of degrees of freedom of the $\chi$ particle, and $\langle \sigma_{\rm ann} v_{\rm rel}
\rangle$ is the thermal average of the annihilation cross section, where $v_{\rm rel}$ refers to the relative velocity of the dark matter particle during freeze-out.  The current value for the DM density, coming from global fits of
cosmological parameters to a variety of observations,  is $\Omega_\chi h^2 = 0.112 \pm 0.006$ \cite{pdg}.

\begin{figure}[t]
\begin{center}
    \includegraphics[width=0.8\textwidth]{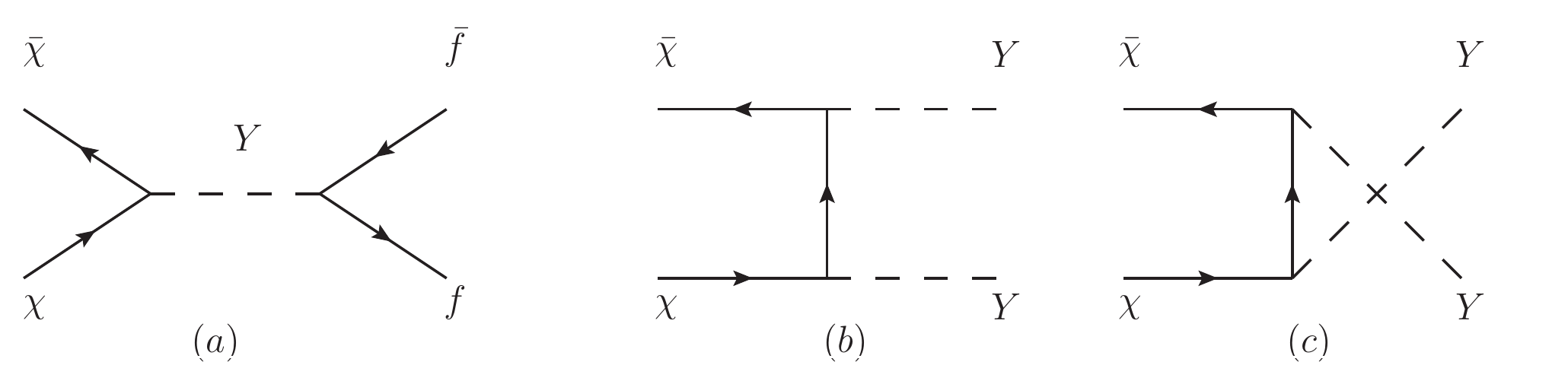}
    \caption{Tree level dark matter annihilation processes inferred from Lagrangian in Eq.~(\ref{eq:L}).}
\label{fig:chichi2yy}
\end{center}
\end{figure}

\begin{figure}[tbhp]
\begin{center}
\includegraphics[width=0.70\textwidth]{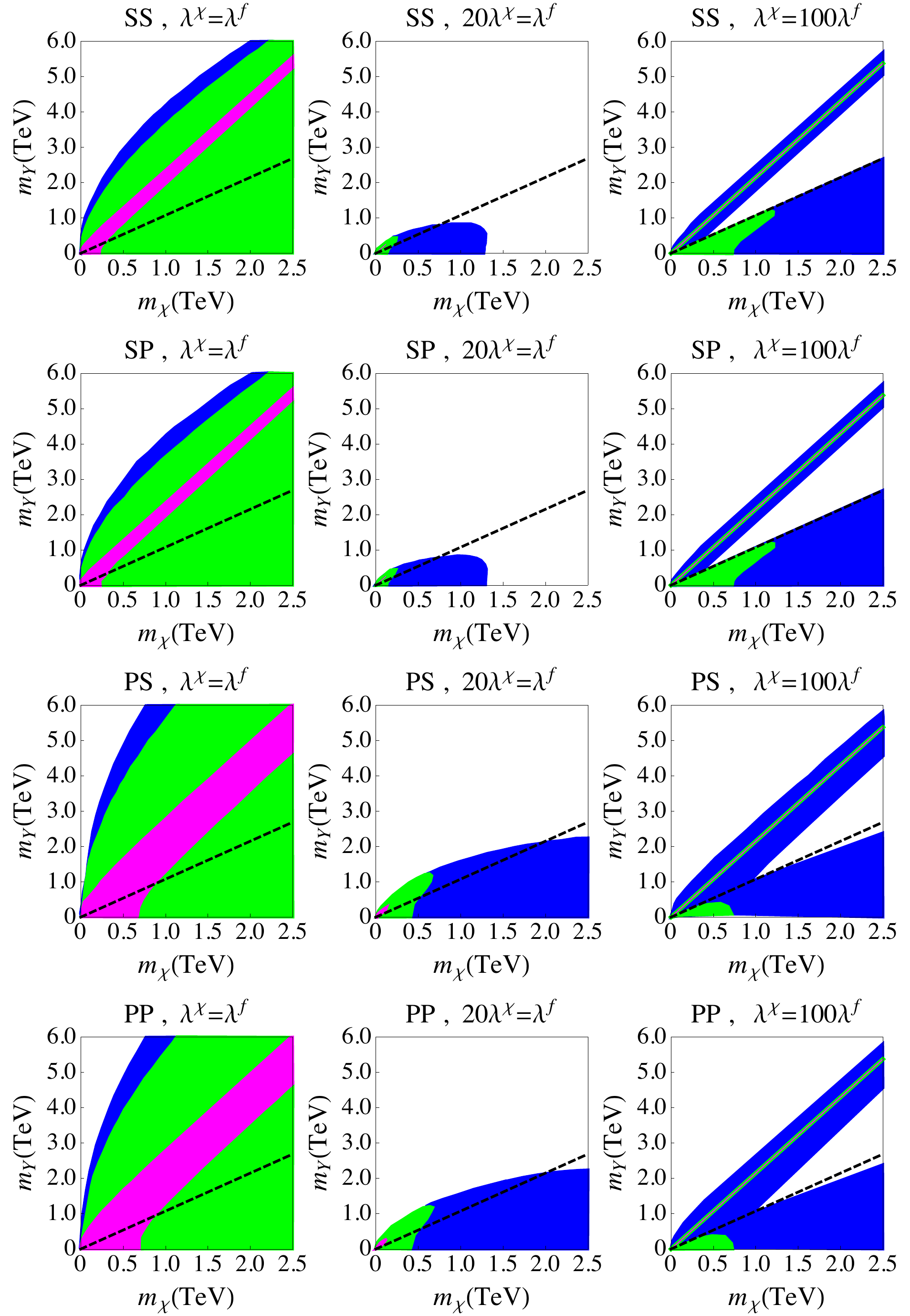}
\caption{ Allowed parameter spaces of $m_\chi$ and $m_Y$ in reproducing the thermal abundance, $\langle\sigma v_{\rm rel}\rangle
\sim  1$ ${\rm pb}\cdot c$. The magenta, green, and blue regions correspond to $0<{\rm Max}[\lambda^\chi, \lambda^f] <0.3$,
$0.3<{\rm Max}[\lambda^\chi, \lambda^f] <1$, and $1<{\rm Max}[\lambda^\chi, \lambda^f] <3$, respectively. From left to right,  (i) $
\lambda_{s,p}^\chi = \lambda_{s,p}^f$ for which all channels are equally important,    (ii) $20\lambda_{s,p}^\chi =
\lambda_{s,p}^f$ for which the $s$-channel dominates over the $t$- and $u$-channels, and (iii) $\lambda_{s,p}^\chi = 100\lambda_{s,p}^f$ for which the $t$- and $u$-channels predominate the contributions.  The shaded region below the dashed line receives contributions from all channels, while above the dashed line only the $s$-channel contributes.}
\label{fig:result1}
\end{center}
\end{figure}

 The abundance of the DM is determined by the $s$-channel annihilation into a SM quark-pair through the exchange of the scalar $Y$, and,
when $E >m_Y$, by $t$- and $u$-channel annihilations into two $Y$ particles with $\chi$ mediated, where $E$ is the energy of the dark
matter particle. The processes are shown in Fig. \ref{fig:chichi2yy}, and the annihilation cross sections are listed in the
following,
\begin{eqnarray}\label{eq:relic}
\langle \sigma v_{\rm rel} \rangle^{SS}&=& N_c\frac{{\lambda_s^\chi}^2 {\lambda_s^f}^2}{2\pi}
  \frac{p_\chi^2  p_f^3}{E^3[(4E^2-m_Y^2)^2 +m_Y ^2 \Gamma_{SS,Y}^2] }
     +\frac{{\lambda^\chi_s}^4}{2\pi}
     \frac{E m_\chi^2  p_\chi^2  p_Y \,\theta(E -m_Y)}{[4(E^2-m_Y^2)m_\chi^2+m_{Y}^4]^2}  \,,
     \nonumber\\
\langle \sigma v_{\rm rel} \rangle^{SP}&=&
 N_c \frac{{\lambda_s^\chi}^2 {\lambda_p^f}^2}{2\pi}
  \frac{ p_\chi^2  p_f }{E [(4E^2-m_Y^2)^2 +m_Y ^2 \Gamma_{SP,Y}^2]}
   +\frac{{\lambda^\chi_s}^4}{2\pi}
   \frac{E  m_\chi^2  p_\chi^2   p_Y \,\theta(E -m_Y) }{[4(E^2-m_Y^2)m_\chi^2+m_Y^4]^2} \,,
   \nonumber\\
\langle \sigma v_{\rm rel} \rangle^{PS}&=&
  N_c \frac{{\lambda_p^\chi}^2 {\lambda_s^f}^2}{2\pi}
    \frac{ p_f^3}{E [(4E^2-m_Y^2)^2 + m_Y ^2 \Gamma_{PS,Y}^2]}
    +\frac{{\lambda^\chi_p}^4}{2\pi}
      \frac{m_\chi^2  p_\chi^2 p_Y^5  \,\theta(E -m_Y) }{E^3 [4(E^2-m_Y^2)m_\chi^2+m_Y^4]^2} \,,
          \nonumber\\
\langle \sigma v_{\rm rel} \rangle^{PP}&=&
 N_c \frac{{\lambda_p^\chi}^2 {\lambda_p^f}^2}{2\pi}
  \frac{E   p_f}{[(4E^2-m_Y^2)^2 +m_Y ^2 \Gamma_{PP,Y}^2]}
  +\frac{{\lambda^\chi_p}^4}{2\pi}
      \frac{m_\chi^2  p_\chi^2 p_Y^5 \,\theta(E -m_Y) }{E^3 [4(E^2-m_Y^2)m_\chi^2+m_Y^4]^2} \,,
\end{eqnarray}
where all flavors will be added on the right hand side, and $p_\chi, p_Y$, and $p_f$ are the
momenta of the $\chi, Y$ and quark, respectively.  Here $N_c=3$ is the number of the quark's colors.  The decay widths of the $Y$ are
given by
\begin{eqnarray}
&& \Gamma_{SS,Y} = \Gamma (Y\to \bar{\chi} \chi)_S + \Gamma (Y\to \bar{f} f )_S    \,, \nonumber\\
&&  \Gamma_{SP,Y} =\Gamma (Y\to \bar{\chi} \chi)_S + \Gamma (Y\to \bar{f} f )_P  \,, \nonumber\\
&& \Gamma_{PS,Y} = \Gamma (Y\to \bar{\chi} \chi)_P + \Gamma (Y\to \bar{f} f )_S  \,, \nonumber\\
&&  \Gamma_{PP,Y} =\Gamma (Y\to \bar{\chi} \chi)_P + \Gamma (Y\to \bar{f} f )_P   \,,
\end{eqnarray}
with
\begin{eqnarray}
&& \Gamma (Y\to \bar{\chi} \chi)_S =  \frac{1}{8\pi} m_Y {\lambda_s^\chi}^2 \beta_\chi^3
\, \theta(m_Y-2m_\chi) \,, \nonumber\\
&& \Gamma (Y\to \bar{\chi} \chi)_P =  \frac{1}{8\pi} m_Y {\lambda_p^\chi}^2 \beta_\chi
\, \theta(m_Y-2m_\chi)  \,, \nonumber\\
 && \Gamma (Y\to \bar{f} f )_S = \frac{N_c}{8\pi} m_Y {\lambda_s^f}^2 \beta_f^3 \,, \nonumber\\
 &&  \Gamma (Y\to \bar{f} f )_P = \frac{N_c}{8\pi} m_Y {\lambda_p^f}^2 \beta_f \,,
\end{eqnarray}
where $\beta_f =\sqrt{1 -4 m_f^2/m_Y^2}$ and $\beta_\chi =\sqrt{1 -4 m_\chi^2/m_Y^2}$.
Freeze-out happens at temperature $T_f\sim m_\chi/20$, and thus the DM energy is around $E\sim m_\chi+ 3m_\chi/40$ (with $T_f\sim m_
\chi v^2 /3$). In Eq.~(\ref{eq:relic}), we have shown four special scenarios denoted by  the superscript ``$SS$", ``$SP$", ``$PS$", and ``$PP$",  for which only the couplings: (i) $\lambda_s^\chi$ and $\lambda_s^f$, (ii) $\lambda_s^\chi$ and $\lambda_p^f$, (iii) $\lambda_p^\chi$
and $\lambda_s^f$, and (iv) $\lambda_p^\chi$ and $\lambda_p^f$ are turned on, respectively.   The scenario for $SS$ is the parity conserving interactions, while the other three are parity violating interactions  \footnote{If the mediator $Y$ is a pseudoscalar particle, then the $PP$ scenario is parity conserving.}.
The present relic can be approximated by
\begin{eqnarray}
 \Omega_\chi h^2 \simeq \frac{0.1~ {\rm pb}\cdot c}{\langle \sigma_{\rm ann} v_{\rm rel}\rangle},
  \end{eqnarray}
 so that it should be $\langle \sigma_{\rm ann} v_{\rm rel}\rangle \sim 1~{\rm pb}\cdot c$. Because $E\simeq p_f, \Gamma_{SS,Y}\simeq \Gamma_{SP,Y}$, and $\Gamma_{PS,Y}\simeq \Gamma_{PP,Y}$, we therefore have  $\langle \sigma v_{\rm rel} \rangle^{SS}\simeq \langle \sigma v_{\rm rel} \rangle^{SP}$ and $\langle \sigma v_{\rm rel} \rangle^{PS}\simeq \langle \sigma v_{\rm rel} \rangle^{PP}$, as shown in Fig.  \ref{fig:result1}.

For the individual scenario given in Eq.~(\ref{eq:relic}), we shall consider three limits: (1) $\lambda_{s,p}^\chi = \lambda_{s,p}^f$ for which the
$t$- and $u$-channels are not negligible compared to the $s$-channel diagram,   (2) $20\lambda_{s,p}^\chi = \lambda_{s,p}^f$ for which the
$s$-channel dominates over the $t$- and $u$-channels, and (3) $\lambda_{s,p}^\chi = 100\lambda_{s,p}^f$ for which the $t$- and $u$-
channels predominate the contributions. To ensure that the calculation can be performed perturbatively, we impose  the absolute value of $
\lambda^\chi$ and $\lambda^f$ to be smaller than 3. We shall consider the three conditions: (i) $0<{\rm Max}[ \lambda^\chi, \lambda^f]
<0.3$, which is denoted by the magenta region, (ii) $0.3<{\rm Max}[\lambda^\chi, \lambda^f] <1$, which is given by the green region,
and (iii) $1< {\rm Max}[\lambda^\chi, \lambda^f] <3$, which corresponds to the blue region. We also assume the universal  $
\lambda_{s,p}^f$, i.e, its value is independent of the flavor of the quark. It should be noted that  for the (blue) region with $\lambda^{\chi,f}>1$
the width of the scalar mediator $Y$ becomes comparable with its mass, so that $Y$ is no longer a good resonant state.

Under these scenarios, we show the allowed parameter space of $(m_\chi, m_Y)$ in Fig. \ref{fig:result1}. Four remarks are in order. First, while maintaining the correct relic abundance,  the region, where the contribution of the $s$-channel resonance is dominant, can satisfy small couplings, $\lambda^\chi = \lambda^f$ and $0< \lambda^{\chi (f)}  <0.3$. Second, the shaded
region below the dashed line corresponding to $E=m_Y$ receives contributions from all channels, while above the dashed line,  where $E
(\approx m_\chi +3 m_\chi/40) < m_Y$, only the $s$-channel contributes and the $t$- and $u$-channels are not kinematically allowed.
Third, for $\lambda^\chi \ll \lambda^f<1 $, the $s$-channel will be predominant; moreover,  due to the nonzero width of the $Y$, the $s
$-channel resonance region with a larger value of $m_Y$ is excluded, so that the allowed parameter space  corresponds to the lighter mediator and lighter $m_\chi$ is preferred in $SS$ and $SP$ scenarios.  See also the central panels in Fig. \ref{fig:result1}. Forth, for $\lambda^\chi \gg \lambda^f$, the $u$- and $t$-channels are predominant in the region with
$m_Y < m_\chi$, whereas the resonant region (with $2E\sim m_Y$), receiving only the $s$-channel contribution, is still allowed.   See also the right panels in Fig. \ref{fig:result1}.

\subsection{Dark matter constraints from the direct detection and correct relic density} \label{sec:direct-detection}

The DM-nucleon interaction occurs via the exchange of scalar particle $Y$ between the dark matter $\chi$ and the nucleon $N$ in the $t$-
channel process \,$\chi N\to \chi N$. For  the $SS$ and $PS$ scenarios, where $\lambda_s^f\not=0$ and $\lambda_p^f =0$, the interactions are spin-independent (SI) on the nucleus side, while for $SP$ and $PP$ scenarios, where $\lambda_s^f=0$ and $\lambda_p^f \not=0$, they are spin-dependent (SD).  The elastic cross section can be expressed as
\begin{eqnarray}\label{eq:elastic}
\sigma_{el}^{SS}(\chi N \to \chi N)&=&\frac{{\lambda_s^\chi}^2 {\lambda_s^f}^2}{\pi} \frac{m_\chi^2 m_N^2}{(m_\chi+m_N)^2 m_Y^4}
f_N^2\,, \nonumber\\
\sigma_{el}^{SP}(\chi N \to \chi N)&=&\frac{{\lambda_s^\chi}^2 {\lambda_p^f}^2}{\pi} \frac{m_\chi^2 p^2}{2(m_\chi+m_N)^2 m_Y^4}
g_N^2 \,,\nonumber\\
\sigma_{el}^{PS}(\chi N \to \chi N)&=&\frac{{\lambda_p^\chi}^2 {\lambda_s^f}^2}{\pi} \frac{p^2 m_N^2}{2(m_\chi+m_N)^2 m_Y^4}
f_N^2\,,\nonumber\\
\sigma_{el}^{PP}(\chi N \to \chi N)&=&\frac{{\lambda_p^\chi}^2 {\lambda_p^f}^2}{\pi} \frac{p^4}{3(m_\chi+m_N)^2 m_Y^4} g_N^2\,,
\end{eqnarray}
where $p=\mu v$ with $\mu\equiv m_\chi m_N/(m_\chi+m_N)$ and  $v\sim 10^{-3} c$ being the present velocity of the DM in the galactic halo. Here the effective coupling $f_N$ of the nuclear matrix elements induced by the scalar SI coupling to quarks is given by \begin{equation}
f_N=\sum_{q=u,d,s} f_{T_q}^{(N)} \frac{m_N}{m_q} + \frac{2}{27} f_{T_G}^{(N)}\sum_{Q=c,b,t} \frac{m_N}{m_Q} \,,\end{equation}
with $f_{T_q}^{(N)}$ and $f_{T_G}^{(N)}$ being hadronic matrix elements \cite{Bottino:2001dj,Ellis:2005mb,Freytsis:2010ne,Gondolo:2004sc},
defined by $\langle N| m_q \bar{q} q |N\rangle=m_N f_{T_q}^{(N)} \bar{u}_N u_N$ and $f_{T_G}^{(N)}=1-\sum_{q=u,d,s} f_{T_q}^{(N)}$, where  $u_N$ is the spinor for the nucleon $N$ and is normalized according to the convention  $\bar{u}_N u_N=2 m_N$.
%Note that the state is normalized according to the non-relativistic convention,
%\begin{eqnarray}
%\langle \tilde{N} ({\bold q})| \tilde{N} ({\bold p}) \rangle=\delta^{(3)} ({\bold q}-{\bold p})\,,
%\end{eqnarray}
 We adopt the
values used by the DarkSUSY package \cite{Gondolo:2004sc},
\begin{eqnarray}
&& f_{T_u}^{(p)}=0.023, \quad  f_{T_d}^{(p)}=0.034, \quad f_{T_s}^{(p)}=0.14, \quad f_{T_G}^{(p)} =0.803,
\nonumber\\
&& f_{T_u}^{(n)}=0.019, \quad  f_{T_d}^{(n)}=0.041, \quad f_{T_s}^{(n)}=0.14, \quad f_{T_G}^{(n)} =0.8.
\end{eqnarray}
The DarkSUSY result is $f_N \simeq 17.5$ compared with $f_N \simeq 14.5$ in \cite{Ellis:2000ds} and $f_N \simeq 12.0$ with a smaller $f_{T_s}^{(N)}=0.053$ in \cite{Cheng:2012qr}. The recent lattice calculation favors a even smaller strangeness content of the nucleon with $f_{T_s}^{(N)}=0.012$, and predicts $f_N \approx 11.5$ \cite{Bali:2012rs}. (See also Ref. \cite{Thomas:2012tg}.) Note that in our case  $f_N$ is not so sensitive to the value of  $f_{T_s}^{(N)}$, compared with the Higgs portal case where the coupling is proportional to the mass of the quark. Basically, the value of $f_N$ in the DarkSUSY is the biggest among these studies. If instead using the lattice value for $f_N$ in the calculation, it is equivalent to adopt the DarkSUSY result but with a 1.5 times smaller value of the product of the couplings, $\lambda^\chi \lambda^f$.

 For obtaining the effective coupling $g_N$ of the nuclear matrix elements induced by the pseudoscalar SD coupling to quarks, we can perform the substitution \cite{Freytsis:2010ne},
\begin{eqnarray}
\langle N'|\bar{q} i \gamma_5 q |N\rangle  \simeq (1 -\eta \delta_q )\Delta q^{(N)} \frac{m_N} {m_q} \bar{u}'_N i \gamma_5 u_N \,,
\end{eqnarray}
where $\delta_u=1, \delta_d=z$,  $\delta_s=w$, and $\eta=(1+z+w)^{-1}$, with $z=m_u/m_d$ and $w=m_u/m_s$. Here $\Delta q^{(N)}$ measures the fraction of the spin carried by the quarks, $q$ and $\bar{q}$, in the nucleon, $N$.
Thus $g_N$ is given by \cite{Cheng:1988im}
\begin{equation}
g_N= (1-\eta)\Delta u^{(N)} \frac{m_N}{m_u} + (1-\eta z)\Delta d^{(N)} \frac{m_N}{m_d} + (1-\eta w)\Delta s^{(N)} \frac{m_N}{m_s}  \,.\end{equation}
Again, we adopt the DarkSUSY numbers,
\begin{eqnarray}
&&\Delta u^{(p)} = \Delta d^{(n)} =0.77 \,,\nonumber\\
&&\Delta d^{(p)} = \Delta u^{(n)} =-0.40 \,,\nonumber\\
&&\Delta s^{(p)} = \Delta s^{(n)} =-0.12 \,,
\end{eqnarray}
and then have $g_p=50.7 $ and $g_n=46.8$, with the same sign\footnote{Using the different formula given in \cite{Cheng:2012qr},  $g_p$ and $g_n$ will have opposite signs, so that the resultant XENON100 exclusion limit becomes even weaker in constraining the $SP$ and $PP$ scenarios.}, compared with  $g_p=64.8 $ and $g_n=61.1$ from the lattice results for quark spin components  \cite{Bali:2012rs}. As $O_{\rm SI} \sim (\bar{\chi} \chi) (\bar{q} q)$ and $O_{\rm SD} \sim (\bar{\chi} \gamma_\mu\gamma_5 \chi) (\bar{q} \gamma^\mu\gamma_5 q)$ operators dominate the interaction of DM with nuclear targets, the direct detection measurements quote the results as bounds on SI and SD cross sections per nucleon, respectively.  However, it is interesting to note that numerically we have $\sigma_{el}(\chi p\to \chi p)\simeq \sigma_{el}(\chi n\to \chi n)$ for the SI and SD operators  considered in this paper. 

\begin{figure}[htb!]
\begin{center}
\includegraphics[width=0.80\textwidth]{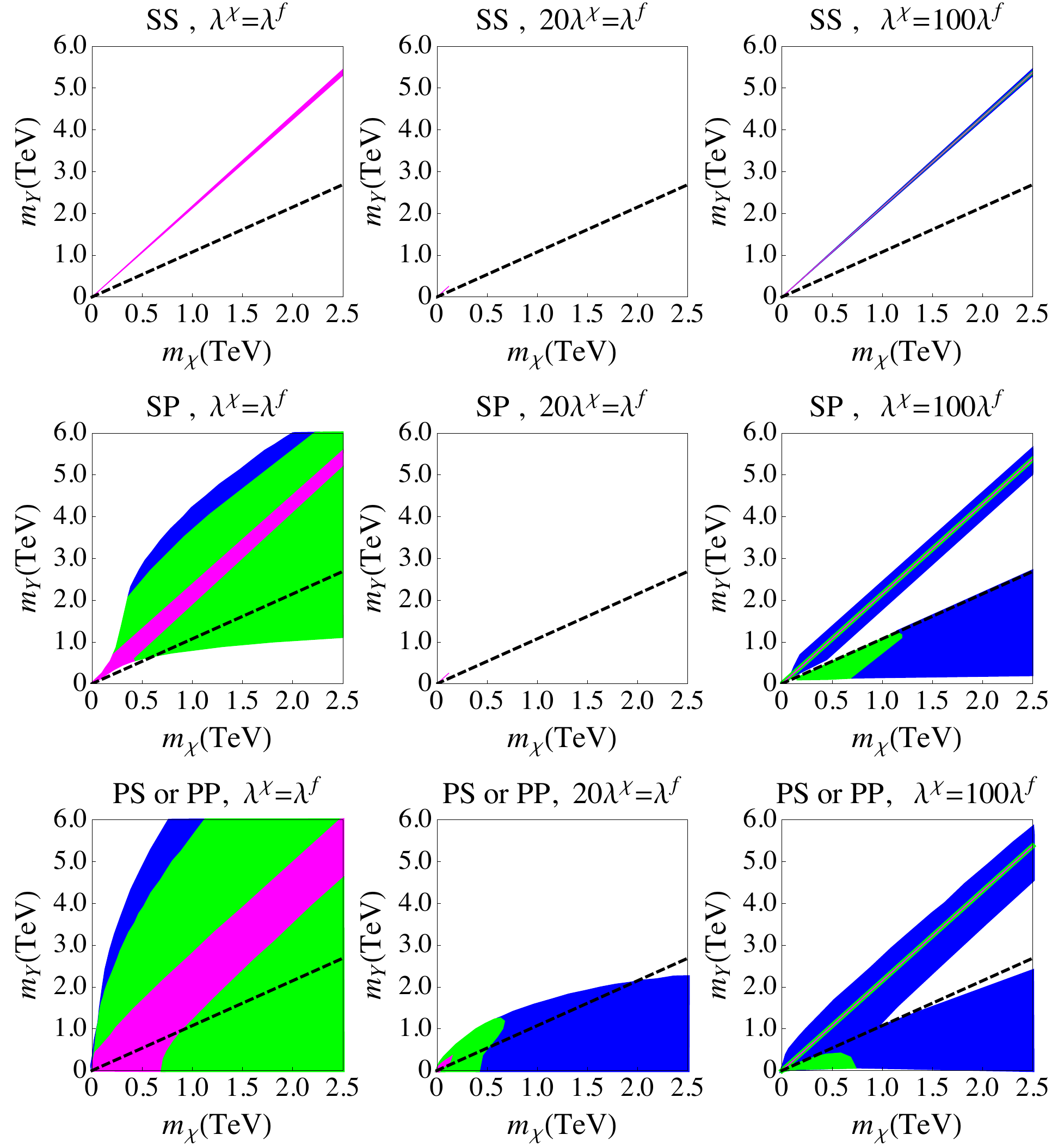}
\caption{Same as Fig. \ref{fig:result1} except for requiring that the correct relic abundance is obtained and the DM-nucleon cross section $\sigma_{el}$ is less than the XENON100 upper limit. The XENON100 exclusion limit has been extended
to the region with $m_\chi>1$~TeV, assuming a linear fit in $m_\chi$.}
\label{fig:sigma_SI}
\end{center}
\end{figure}

In the direct detection, compared with $\sigma_{el}^{SS}$,  one has that $\sigma_{el}^{SP}$ and $
\sigma_{el}^{PS}$ are velocity suppressed by $(v/c)^2 m_\chi^2 /(m_\chi+m_N)^2 $ and  $(v/c)^2 m_N^2 /(m_\chi+m_N)^2 $, respectively, while $\sigma_{el}^{PP}$ is further
suppressed by  $(v/c)^4 m_N^2 m_\chi^2 /(m_\chi+m_N)^4 $. As for $m_\chi \gg m_N$, we have  $\sigma_{el}^{SS} \gg \sigma_{el}^{SP}\gg\sigma_{el}^{PS}\gg \sigma_{el}^{PP}$. These results are indicated in Fig. \ref{fig:sigma_SI}.

\begin{figure}[htbp]
\begin{center}
\includegraphics[width=0.85\textwidth]{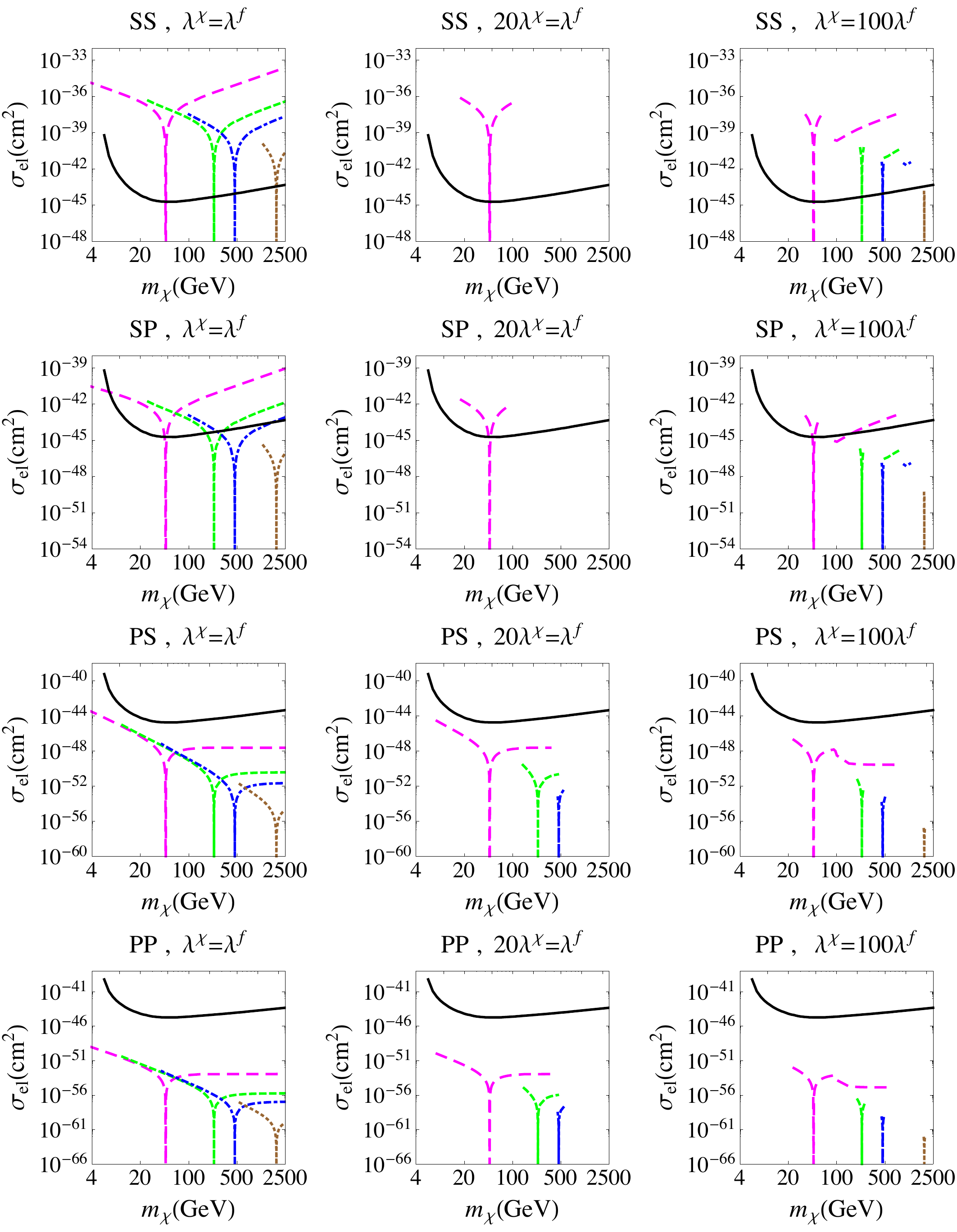}
\caption{DM-nucleon scattering cross section as a function of $m_\chi$, predicted by requiring the
values of couplings to satisfy the correct relic constraint, where we have set the bound of the couplings to be $ {\rm Max}
[\lambda^\chi, \lambda^f] <1$. The XENON100  experimental 90\%-C.L. upper limit is depicted by the solid curve.   From left to right,
the long dashed, short dashed, dot-dashed, dotted curves respectively correspond to the following values, $m_Y=100, 500, 1000$, and 4000
GeV, as input.}
\label{fig:sigma-mchi}
\end{center}
\end{figure}

Requiring that the correct relic density is obtained by the thermal freeze-out, and the DM-nucleus elastic cross section is consistent with the
current 90\%-C.L. upper bound by the  XENON100 measurement \cite{Aprile:2012nq}, we depict the allowed parameter space of $(m_\chi, m_Y)$ in Fig. \ref{fig:sigma_SI}.  For  the ``$SS$"  scenario, in addition to the condition $\lambda^f \gg \lambda^\chi$ for which the $s$-channel is
predominant in the thermal relic density and few solutions exist,  only the region that is very close to  the $s$-channel resonance of
annihilation in Eq. (\ref{eq:relic}) is allowed.  (See also Fig.~\ref{fig:sigma-mchi}.) If the Higgs-portal models of DM generate the operator as
our ``$SS$" scenario, the allowed solution will be very close to the $s$-channel resonance of annihilation, so that $m_\chi\approx m_h/2$ \cite{LopezHonorez:2012kv}. This is also called the  "{\it resonant Higgs portal}"  model \cite{LopezHonorez:2012kv}.
The result will be further constrained  by the invisible Higgs decay branching ratio  ${\rm Br}_{inv}$, for which  ${\rm Br}_{inv} \lesssim 20\%$  if the Higgs production rate at the LHC  \cite{ATLAS:2012ae,Chatrchyan:2012tx,:2012gk,:2012gu} arises from the SM Higgs, or ${\rm Br}_{inv} \lesssim 65\%$ if new contributions exist \cite{Carmi:2012in}. As for other scenarios, compared with the constraint from the correct relic density, the direct detections  offer marginal constraints on $SP, PS$ and $PP$ interactions.  The $PS$ scenario is related to the "{\it pseudoscalar Higgs portal}" model \cite{LopezHonorez:2012kv}.

Taking  the thermal relic constraint into account,  in Fig. \ref{fig:sigma-mchi} we compare the predictive curves of the DM-nucleon scattering cross section with the current XENON100 upper bound \cite{Aprile:2012nq}. Most $(m_\chi, m_Y)$ region is excluded  (except the resonant part) in the $SS$ scenario as well as in the $SP$ scenario with $\lambda^\chi \ll \lambda^f$. However, a large region is still viable in $SP$ (with $\lambda^\chi=\lambda^f$ or $\lambda^\chi \gg \lambda^f$), $PS$ and $PP$ scenarios.  For $PS$ and $PP$ scenarios, the predicted DM-nucleon cross section is fully below the XENON100 bound and the main constraint comes from the thermal relic abundance. The results can be easily seen from Fig. \ref{fig:sigma-mchi} and are due to the fact that  $\sigma_{el}^{SP}, \sigma_{el}^{PS}$ and $\sigma_{el}^{PP}$ are velocity suppressed,  compared with $\sigma_{el}^{SS}$. Meanwhile,
if the DM annihilation cross section is dominated by the $t$- and $u$-channel amplitudes, i.e., $\lambda^\chi \gg \lambda^f$, the solutions are mainly located in the region where $m_\chi >m_Y$, in addition to the narrow region close to  the $s$-channel resonance where the couplings need to be small enough to maintain the correct relic density.

Before proceeding, two remarks are in order. (i) As shown in Fig.~\ref{fig:sigma-mchi}, for a fixed $m_Y$ with $m_\chi \gg m_Y /2$ and for $\lambda^\chi \simeq \lambda^f$,  $\sigma_{el}$ increases in $SS$ and $SP$ scenarios as $m_\chi$ becomes much larger, while $\sigma_{el}$ is independent of $m_\chi$ in $PS$ and $PP$ scenarios. (ii) For a fixed $m_\chi$ with $m_\chi \gg m_Y /2$ and for $\lambda^\chi \simeq \lambda^f$, if choosing a larger $m_Y$ as the input, it
will result in a smaller $\sigma_{el}$.  These two properties are be understood as follows. In the large $m_\chi$ limit, because $\langle \sigma_{\rm ann} v_{\rm rel}\rangle
\sim 1~{\rm pb}\cdot c$~ (=const.) $\propto  (\lambda^\chi \lambda^f)^2/m_\chi^2$, we therefore have the DM-nucleon scattering
cross section to be $\sigma_{el}\propto (\lambda^\chi \lambda^f)^2/m_Y^4 \propto m_\chi^2/m_Y^4 $ for $SS$ and $SP$ scenarios, and  $\sigma_{el}\propto (\lambda^\chi \lambda^f)^2/(m_\chi^2 m_Y^4) \propto 1/m_Y^4 $ for $PS$ and $PP$ scenarios.

\subsection{Dark matter constraints from the LHC as well as the correct relic density and direct detection}\label{sec:LHC}

In addition to the DM mass constraints due to the correct relic density and the direct detection, we will further consider the collider bound
due to the monojet measurements with  large missing transverse energy ($\not\!\! E_T$) in the final states, and dijet resonance search.

\subsubsection{Monojet + $\not\!\! E_T$ search}
At the hadron colliders, the  dark matter particles can be produced and subsequently escape from the detector since they weakly interact
with SM particles. Thus the signature of dark matter is a process with some missing energy. We can therefore use the monojet searches to
constrain our dark matter model here.

\begin{figure}[t]
\begin{center}
\includegraphics[width=0.3\textwidth]{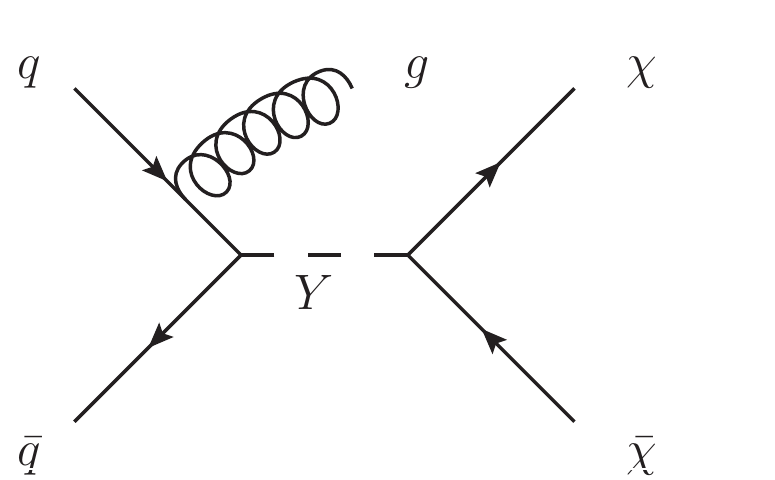}
\caption{Dark sector productions through processes involving  a single jet, where the jet can be one of quark, antiquark, and gluon, and the
remaining two are from different accelerated proton beams at the LHC.}
\label{fig:qqrY}
\end{center}
\end{figure}

In experiments, monjet + $\not\!\! E_T$ final states have been studied by CDF \cite{Aaltonen:2008hh}, CMS \cite{Chatrchyan:2011nd}, and
ATLAS \cite{Aad:2011xw,Martinez:2012ie}.  Recently, ATLAS \cite{Martinez:2012ie} has  analyzed monojets with varying jet $p_T$ cuts using
1.00 fb$^{-1}$ of data, for which the number of the observed monjet events is in agreement with the Standard Model predictions.  As studied in \cite{lhc-4,An:2012va},  the ATLAS very high $p_T$ (veryHighPt) analysis can obtain the optimal  bound for interactions involving the dark sector except for that the CDF data provide a little more stringent bound than LHC data for $m_\chi\lesssim$ 25 GeV \cite{Shoemaker:2011vi}. The ATLAS search in an event sample of the veryHighPt region has given 95\% CL  cross section upper limits of 0.045 pb  for new phenomena \cite{Martinez:2012ie}.  The leading channels containing the dark matters in the final state with the monojet plus missing transverse energy
search at hadron colliders are shown in Fig. \ref{fig:qqrY}, while the Standard Model background consists mainly of $(W\to \ell^{\rm inv}
\nu)+ {\rm jet}$ and $(Z\to \nu \bar\nu) + {\rm jet}$ final states, where $\ell^{\rm inv}$ denotes for the charged lepton lost by the
detector.  As the DM mass increases, the resulting production cross section decreases due to the smaller phase space, and therefore we can expect that the monojet constraint becomes weaker.

Together with the constraints from the correct relic density and the exclusion bound of the XENON100 direct detection measurement, we will
use the recent ATLAS results on the monojet search to  find the allowed parameter space  of $(m_Y, m_\chi)$.  We have used the
MadGraph 5 \cite{Alwall:2011uj} and CTEQ6.1L parton distribution functions \cite{Stump:2003yu} to simulate the monojet  plus  missing energy events in the veryHighPt region, where the missing energy is $
\not\!\! E_T > 300$ GeV, and one jet has a large transverse momentum  $p_T(j_1) > 350$ GeV and pseudo-rapidity $|\eta(j_1)|<2$.  The
detailed definition of the veryHighPt region can be found in \cite{Martinez:2012ie}.

\subsubsection{Dijet resonance search}

\begin{figure}[t]
\begin{center}
\includegraphics[width=1.0\textwidth]{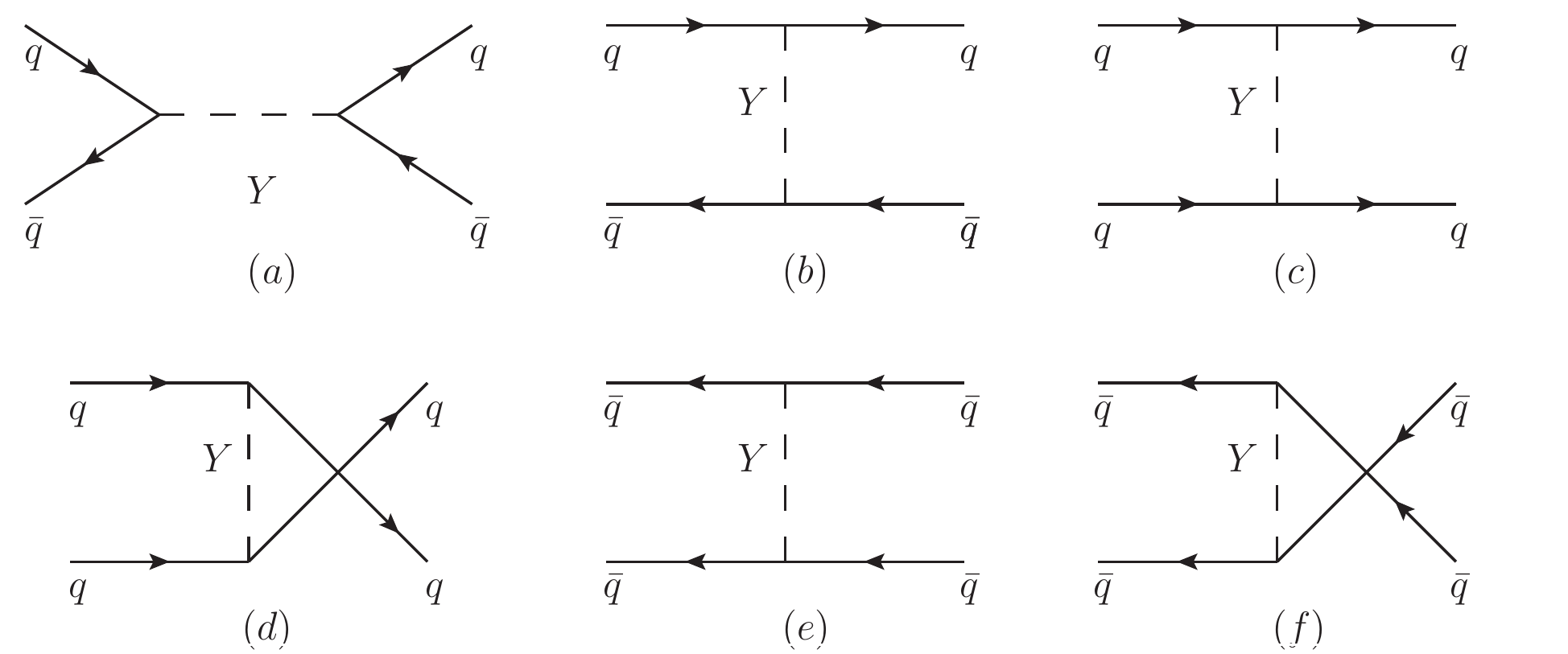}
\caption{Leading dijet processes in the $s$-, $t$- and $u$-channels, where the initial and final states contain two quark jets and the mediator is the $Y$.  (a) is the $s$-channel, (b), (c)  and (e) are $t$-channels, and (d) and (f) are $u$-channels. Dijet experiments do not distinguish between the quark and antiquark.}
\label{fig:FDdijet}
\end{center}
\end{figure}

The hadron colliders have performed the search for new particles beyond the standard model in the dijet mass spectrum.  The process relevant to the dark matter search in the experiments is the $s$-channel production and decay of  the narrow dijet resonance, $Y$,  shown in Fig. \ref{fig:FDdijet}, where the contributions arising from the $t$- and $u$-channels are negligible and will be discussed later. Note that the gluon-gluon fusion into $Y$ through a quark triangle loop is highly suppressed, compared to $q \bar{q}$ annihilation  into $Y$ that we have shown in Fig. \ref{fig:FDdijet}.  CDF collaboration \cite{Aaltonen:2008dn}, using proton-antiproton collision data corresponding an integrated luminosity of 1.13 fb$^{-1}$, has presented a search for new narrow particles whose decays produce dijets with invariant mass in the region $260\ {\rm GeV}<m_Y <1400\ {\rm GeV}$. Recently, CMS and ATLAS collaborations \cite{CMS:2012yf,Aad:2011fq} have published the dijet searches for new narrow resonances in the region $1\ {\rm TeV}<m_Y <4.3\ {\rm TeV}$ and $900\ {\rm GeV}<m_Y <4\ {\rm TeV}$  using integrated luminosities 5 fb$^{-1}$ and 1 fb$^{-1}$, respectively, at a center-mass-energy $\sqrt{s}=7$~TeV. All the results show no evidence of new narrow resonance production over the SM background.

For the dijet resonance search, the LHC data measured by the CMS and ATLAS collaborations give the most stringent constraint in the region $900\ {\rm GeV}<m_Y <4.3\ {\rm TeV}$, while the CDF measurement gives stronger constraint for $m_Y <900 \ {\rm GeV}$. Because the results given by CMS and ATLAS are quite similar, in the numerical analysis we will therefore mainly use the CMS and CDF data.  The contributions of $t$-and $u$-channels in Fig. \ref{fig:FDdijet} are negligible due to the following two reasons.  First, the $t$- and $u$-channel scatterings  mainly contribute to the small angle region, and can be suppressed by the pseudorapidity cut. Second, the bump-like component of the resonances is relevant only to the $s$-channel, and dijet measurement can set  the upper limit on it \cite{Harris:2011bh}.

\subsubsection{Results and discussions}

\begin{figure}[htb]
\begin{center}
\includegraphics[width=0.80\textwidth]{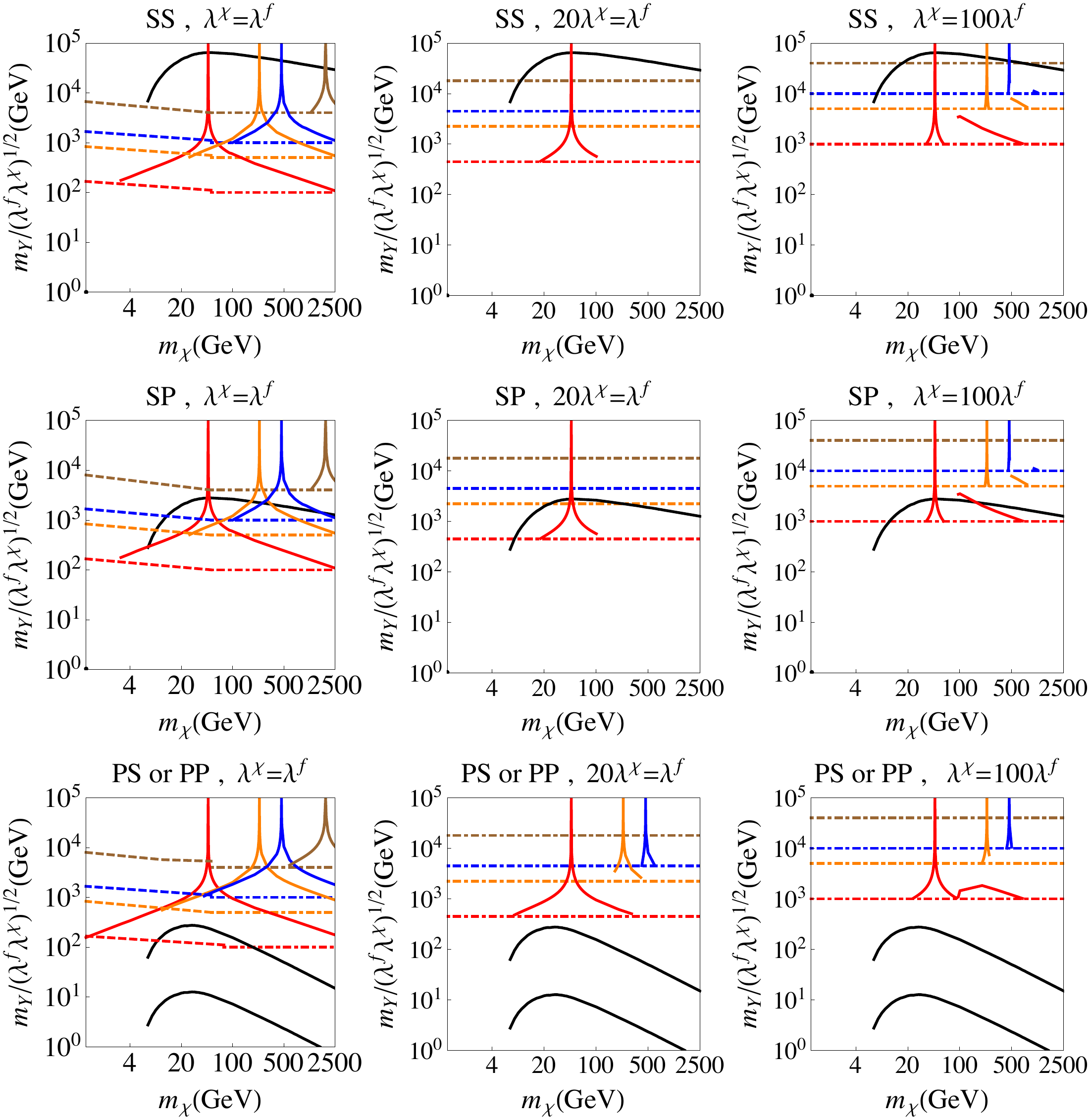}
\caption{$m_Y/(\lambda^\chi \lambda^f)^{1/2}$ vs. $m_\chi$.  (i) The curves with a peak satisfy the correct relic density.  From
left to right, the red, orange, blue, and brown curves correspond to $m_Y=100, 500,1000, 4000$~GeV, respectively. (ii) The lower bound, constrained by the ATLAS null search for monojet + large $\not\!\! E_T$ final states and ${\rm Max}[\lambda^\chi,
\lambda^f] < 1$, is denoted as the dashed or dot-dashed (horizontal) curve, where the former corresponds to the monojet cross section $\simeq 0.045$ pb and the latter is ${\rm Max}[\lambda^\chi, \lambda^f] = 1$. From up to down, the brown, blue, orange, and red curves correspond to $m_Y=4000, 1000, 500, 100$~GeV, respectively.
(iii) The XENON100  experimental 90\%-C.L. upper limit is depicted by the black curve, below which the region is excluded; in the third row, the upper curve is for the $PS$ scenario, and the lower is for the $PP$ one. For curves on the
left hand side of the peak, with $m_Y\gg m_X$ and narrow width limit of the $Y$, the value of $m_Y/(\lambda^\chi \lambda^f)^{1/2}$ is
equivalent to  $\Lambda$.}
\label{fig:sigma-eff}
\end{center}
\end{figure}

\begin{figure}[htb]
\begin{center}
\includegraphics[width=0.75\textwidth]{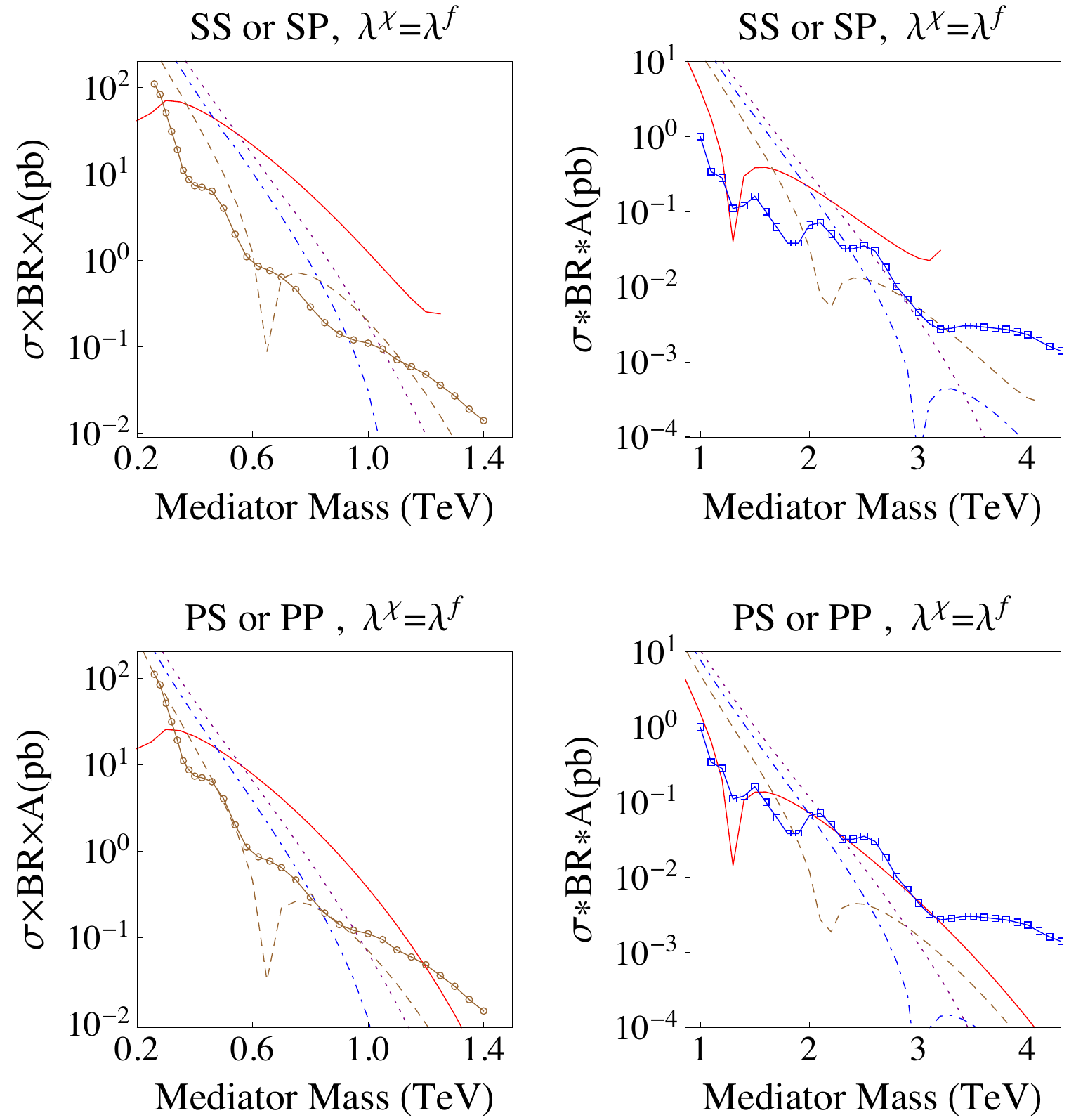}
\caption{The theoretical prediction for $\bar{q} q\to \bar{q} q$ via the scalar resonance $Y$ in colliders as a function of the  mediator (resonance) mass, where the CDF (left panels) and CMS (right panels)  95\% CL upper limits on $\sigma \times B \times A$ for the dijet production of the types $q \bar q$, $\bar{q} \bar{q}$ or $q q$,  are denoted by  open boxes. By requiring the
values of couplings satisfy the correct relic density and narrow width $\Gamma_Y \lesssim 0.1 m_Y$ constraints, the predicted solid, dashed, dash-dotted, and dotted curves (from left to right) respectively correspond to $m_\chi= 100, 300, 500, 700$~GeV in the left panels, and $m_\chi= 500, 1000, 1500, 2000$~GeV in the right panels.}
\label{fig:dijet}
\end{center}
\end{figure}

\begin{figure}[htb]
\begin{center}
\includegraphics[width=0.95\textwidth]{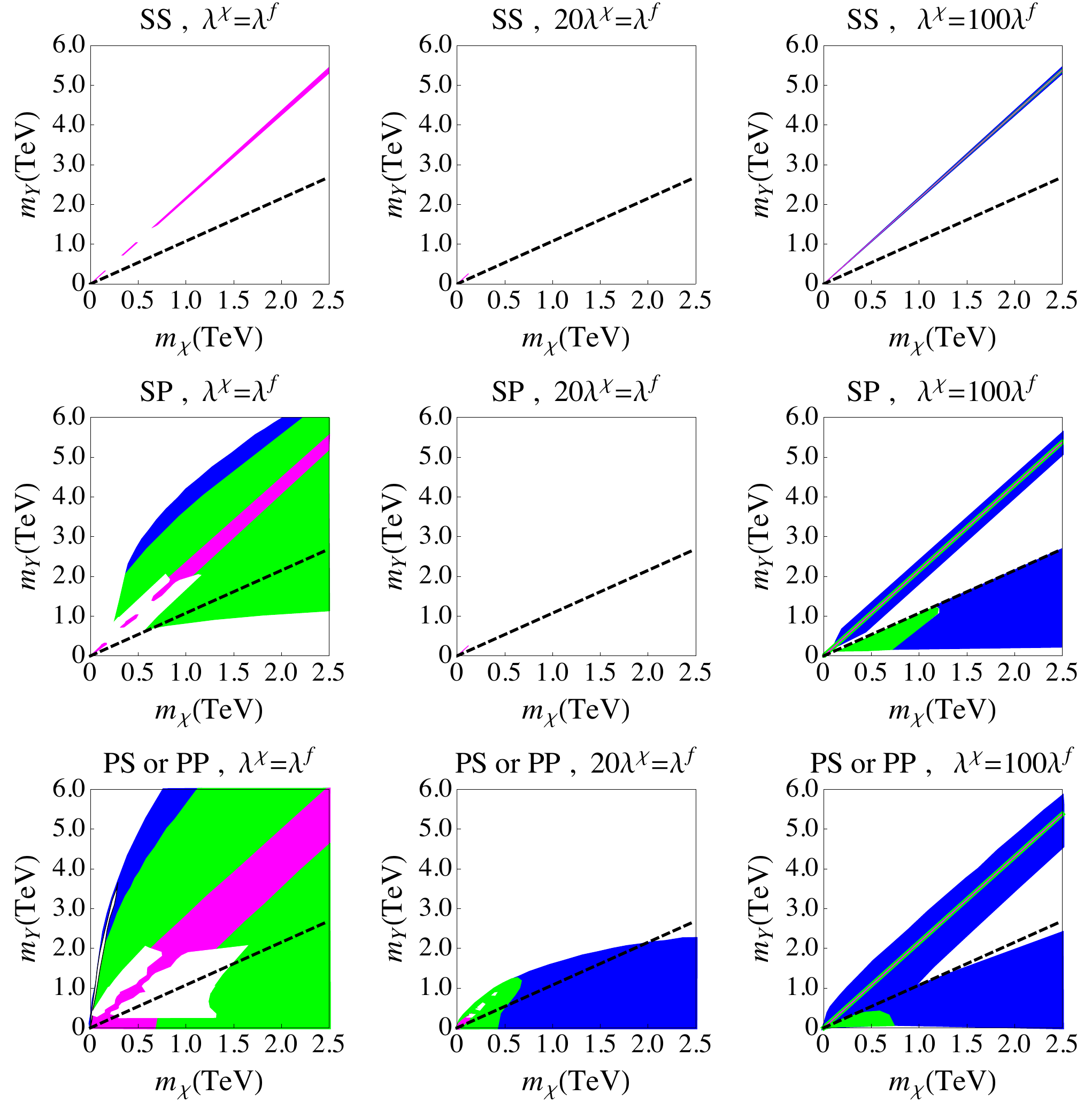}
\caption{Same as Fig.~\ref{fig:sigma_SI} except that (i) the small region with white color but bounded by the black line is excluded by the  ATLAS null search for monojet + large $\not\!\! E_T$ final states, and (ii) the white color region with smaller couplings is excluded by dijet measurements at the colliders. }
\label{fig:LHC}
\end{center}
\end{figure}

  The numerical results are depicted in Figs. \ref{fig:sigma-eff}, \ref{fig:dijet} and \ref{fig:LHC}. The monojet results are summarized as follows. (i) The monojet  constraint is  basically not  stronger than the combination of the correct relic density and direct detection.  As shown in Fig.  \ref{fig:LHC}, for the region satisfying $1<\lambda^\chi (= \lambda^f) <3 $ and only for $PS$ and $PP$ scenarios,  a very small fraction bounded by the black curve can be further excluded by the LHC monojet constraint. Note that,  for the region with $1<\lambda^{\chi (f)} <3$, although the calculation is perturbatively convergent, the decay width of the $Y$ particle may be comparable with its mass, so that $Y$ is no longer a good resonant state. 
 (ii) Because the monojet cross section is proportional to $(\lambda^f \lambda^\chi)^2$, where the coupling is determined by the relic constraint (see also Fig. \ref{fig:result1}),  therefore the monojet exclusive region  for the $SS$ scenario is equal to the one for $SP$, while the monojet exclusive region for $SP$ is equal to the one for $PP$, as depicted in Figs. \ref{fig:sigma-eff} and \ref{fig:LHC}.  (iii) In Fig. \ref{fig:sigma-eff}, we show that, for $ {\rm Max}[\lambda^\chi, \lambda^f] < 1$,  the LHC monojet bound is always weaker than the constraint due to the correct relic density.    (iv) Comparing our direct detection calculations in Eq. (\ref{eq:elastic}) with the results
obtained from the operators defined in the effective theory framework,
  \begin{eqnarray}
  {\cal O} = \frac{(\bar\chi \Gamma_1 \chi)( \bar{q} \Gamma_2  q)}{\Lambda^2} \,,
  \end{eqnarray}
where $\Gamma_{1,2} \equiv \bf{1}$ or $\gamma_5$, we have $m_Y /(\lambda^\chi \lambda^f)^{1/2} \equiv \Lambda$. However, this
relation is no longer correct in the calculations of the relic density and LHC monojet bound. For the relic density, it can be correct only when
taking the limits: $m_\chi \ll m_Y$ and  $\Gamma_Y \ll m_Y$. In the LHC monojet search, because the LHC can not only resolve the
interaction but also produce the on-shell mediator, so that   $m_Y/(\lambda^\chi \lambda_f)^{1/2}$ depends on the value of $m_Y$ as
shown in Fig. \ref{fig:sigma-eff}.

 For the dijet resonance search in the CDF and CMS experiments \cite{Aaltonen:2008dn,CMS:2012yf,Chatrchyan:2011ns,CMS_note}, the best bin width is somewhat larger than the dijet invariant mass resolution $\sigma$, {\it e.g.} in the CMS measurements,
 $\sigma/m_{jj}  = 0.045 + 1.3/  m_{jj}^{1/2}$ with $m_{jj}$ in units of GeV being the dijet mass \cite{CMS_note};
the half-width of the resonance is significantly less than the experimental Gaussian resolution $\sigma$. Conservatively speaking, these experimental results can be used to constrain the mass of the new particle with the narrow width $\Gamma_R \lesssim  0.1 m_R$, where $\Gamma_R$ and $m_R$ are the decay width and mass of the resonance.  In Fig. \ref{fig:dijet}, we show the observed upper limit at the 95\% confidence level on $\sigma \times B \times A$ for $q \bar q$ (or $q q, \bar{q}\bar{q}$) resonance from the inclusive analysis, compared to the predictions for processes with the $s$-, $t$-, and $u$-channel interactions via the scalar resonance, $Y$. We calculate the the results using CTEQ6.1L parton distribution functions \cite{Stump:2003yu}. Here $\sigma$ is the resonance production cross section, $B$ is the branching fraction of the resonance decaying into the jet-jet final state, and $A$ is the acceptance  for the kinematic requirements. For CDF, the acceptance requires that the rapidity of the leading two jets satisfies $|y|<1$. For CMS, the acceptance corresponds to the pseudo-rapidity separation  $\Delta\eta$ of the two leading jets satisfies $|\Delta\eta|<1.3$, and the two jets are also located in $|\eta| <2.5$; $A\approx 0.6$ for isotropic decays independent of the resonance mass.

 Because the dijet cross section is mainly proportional to $(\lambda^f)^4$, where the coupling is determined by the relic constraint  (see also Fig. \ref{fig:result1}),  therefore the dijet exclusive region  for the $SS$ scenario is equal to the one for $SP$, while the monojet exclusive region for $SP$ is equal to the one for $PP$ as depicted in Figs. \ref{fig:dijet} and \ref{fig:LHC}. 
  Comparing the region plot of  $(m_Y$, $m_\chi$) given in  Fig. \ref{fig:LHC} with that in Fig. \ref{fig:sigma_SI}, we find that the dijet measurements offer strong constraint on the cases of $\lambda^\chi =\lambda^f$, where a large portion of the region corresponding to  smaller couplings is further excluded.


\section{Discussions and Conclusions}\label{sec:conclusions}

Before we summarize, let us briefly discuss the constraints from indirect detections and diphoton resonance searches. 

For the indirect detections, dark matter annihilation in the galactic halo can generate observable signals. The annihilation cross section $\langle\sigma v\rangle$, which can also read from Eq. (\ref{eq:relic}), is proportional to $v^2$ for $SS$ and $SP$ scenarios, but is approximately independent of $v$ for $PS$ and $PP$ scenarios, where $v\sim 10^{-3} c$ is the velocity of the DM in the galactic halo. Compared to the DM annihilation cross section at thermal freeze-out, the present $\langle\sigma v\rangle$ in the latter scenarios will be roughly the same as $ 3\times 10^{-26}$ cm$^3$ s$^{-1}$, while the one in the former scenarios is velocity suppressed since the velocity of the DM is $\sim 1/\sqrt{10} c$ at freeze-out temperature. Thus, the FERMI-LAT measurements from observations of Milky Way dwarf spheroidal galaxies \cite{Ackermann:2011wa} and BESS-Polar II \cite{Englert:2011aa} data have no constraints on the $SS$ and $SP$ scenarios due to their velocity suppression, but  disfavor the dark matter mass $m_\chi \lesssim$ 30 GeV for $PS$ and $PP$ scenarios. Nevertheless, one should note that the recent FERMI-LAT search for DM in gamma-ray lines and the inclusive photon spectrum  \cite{Ackermann:2012qk} does not constrain all interacting scenarios that we consider in this paper.

The results for the diphoton resonance analyses at the collider experiments may be relevant to the exotic $Y$ production that we consider in the present work. The diphoton channel has been used to search for the Higgs 
\cite{Hawkings} and Randall-Sundrum graviton \cite{Aad:2012cy}. As for the search for Higgs decaying into diphoton at the LHC and Tevatron, the three main Higgs production channels are the gluon-gluon fusion, vector boson fusion, and vector boson associate production. In our case, the $Y$ particle is produced mainly from the $q \bar{q}$ annihilation.  Requiring that the couplings for the DM into the quark pair are constrained by the correct relic abundance, we find that the cross section for $p p$ into $\gamma \gamma$ via $Y$ is at least 100 times smaller than that via the Higgs. Therefore, the contribution to the diphoton production due to the exotic scalar $Y$ particle can be negligible.  In other words, the diphoton resonance experiments cannot offer the efficient constraint to the mass of the $Y$.

In summary, we have studied the experimental constraints, which are due to measurements from the correct relic abundance, direct/indirect detections, and colliders,  on a scenario that a Dirac fermionic DM  interacts with SM quarks via a scalar mediator in a model-independent way. We respectively consider four special scenarios denoted as $SS, SP, PS$ and $PP$, where the former one is the parity conserving interaction, while the latter three are parity violating. Our present study can apply to the case of the pseudoscalar mediator, for which the $PP$ scenario is the parity conserving interaction. For each  interaction, we take three limits for couplings,  $\lambda_{s,p}^\chi = \lambda_{s,p}^f, \lambda_{s,p}^\chi \ll \lambda_{s,p}^f$ and $\lambda_{s,p}^\chi \gg \lambda_{s,p}^f$,  in the analysis.  The main results are summarized as follows.
\begin{itemize}
\item
Requiring $\lambda^{\chi,f}<1$,  so that the resulting width of the scalar mediator $Y$ is smaller than its mass, the $(m_\chi, m_Y)$ parameter space maintaining the correct relic abundance is very small for the couplings $\lambda_{s,p}^\chi \ll \lambda_{s,p}^f$ or $\lambda_{s,p}^\chi \gg \lambda_{s,p}^f$, as shown in Fig. \ref{fig:result1}. For $\lambda_{s,p}^\chi \gg \lambda_{s,p}^f$, in addition to a narrow region  corresponding to the $s$-channel resonance, the allowed region satisfying $m_Y<m_\chi$ is dominated by the $u$- and $t$-channels of  the relic annihilation processes.

\item
In the direct detection,  the $SS$ and $PS$ interactions, where $\lambda_s^f\not=0$ and $\lambda_p^f =0$, are spin-independent on the nucleus side, while the $SP$ and $PP$ interactions, where $\lambda_s^f=0$ and $\lambda_p^f \not=0$, are spin-dependent.   Compared with the direct detection cross sections $\sigma_{el}^{SS}$,  $\sigma_{el}^{SP}$ and $
\sigma_{el}^{PS}$ are velocity suppressed by $(v/c)^2 m_\chi^2 /(m_\chi+m_N)^2 $ and  $(v/c)^2 m_N^2 /(m_\chi+m_N)^2 $, respectively, where the DM velocity $v\sim 10^{-3}c$, while $\sigma_{el}^{PP}$ is further
suppressed by  $(v/c)^4 m_N^2 m_\chi^2 /(m_\chi+m_N)^4 $. As for $m_\chi \gg m_N$, we have  shown $\sigma_{el}^{SS} \gg \sigma_{el}^{SP}\gg\sigma_{el}^{PS}\gg \sigma_{el}^{PP}$.   Compared with the results constrained by the correct relic abundance in Fig. \ref{fig:result1}, the XENON100 null measurement further excludes most $(m_\chi, m_Y)$ region in the $SS$ scenario (except the resonance region) and in the $SP$ scenario with $\lambda^\chi \ll \lambda^f$.

\item
The current monojet  constraint is  not  stronger than that from the requirement of the correct relic density and the null result by the XENON100 direct detection. Only a very small fraction located in the region with  $1<\lambda^\chi (= \lambda^f) <3 $ can be further excluded in $PS$ and $PP$ scenarios.  However,  although the calculation is perturbatively convergent in that region, the decay
width of the $Y$ particle may be comparable with its mass, so that $Y$ is no longer a good resonant state.

\item
We have used the results from the dijet resonance search in the CDF and CMS experiments to constrain the mass of the scalar mediator $Y$ with the narrow width $\Gamma_Y \lesssim 0.1 m_Y$. Fig. \ref{fig:LHC} has shown the allowed parameter space of $m_\chi$ (DM's mass) and $m_Y$ (mediator's mass), constrained by the correct relic abundance, the null result at the XENON100 direct detection and the bounds due to the collider monojet and dijet searches.  We find that the dijet measurements offer strong constraint on the case of $\lambda^\chi \simeq \lambda^f$, where a large part of the region corresponding to  smaller couplings is further excluded.
\end{itemize}

\acknowledgments \vspace*{-1ex}
This research was supported in part by the National Center for Theoretical Sciences and the
National Science Council of R.O.C. under Grant Nos. NSC99-2112-M-033-005-MY3 and NSC101-2811-M-033-012.

\end{document}